\documentclass[%
preprint,
 amsmath,amssymb,
 aps,
]{revtex4-2}

\usepackage{graphicx}
\usepackage{bm}

\usepackage{caption}
\usepackage{subcaption}
\usepackage{mathtools}
\usepackage{float}
\usepackage{tabularx}
\usepackage{booktabs}

\begin{document}


\title
{Internal Representations in
Spiking Neural Networks, criticality and the Renormalization Group}

\author{João Henrique de Sant'Ana  }
\email{joao.henrique.santana@usp.br, orcid {0000-0003-1348-0974}}
\author{Nestor Caticha }%
 \email{ncaticha@usp.br, orcid {0000-0002-7446-6765}}
\affiliation{ 
 Instituto de Física, Universidade de Sao Paulo, SP, Brazil
}%

\date{\today}

\begin{abstract}
Optimal information processing in peripheral sensory systems has been associated in several examples to the signature of a critical or near critical state. Furthermore, cortical systems have also been described to be in a critical state in both wake and anesthetized experimental models, both {\it in vitro} and {\it in vivo}. We investigate whether  a similar signature characterizes the internal representations (IR) of a  multilayer (deep) spiking artificial neural network performing computationally simple but meaningful  cognitive tasks, using a methodology inspired in the biological setup, with cortical implanted electrodes in rats, either freely behaving  or under different levels of anesthesia. 
The increase of the characteristic time of the decay of the correlation of fluctuations of the IR,  found when the network input changes,  are indications of a broad-tailed distribution of IR fluctuations.  The broad tails are present even when the network is not yet capable of performing the classification tasks, either due to partial training or to the effect of a low dose of anesthesia in a simple model.  However,  we don´t find enough evidence of power law  distributions of avalanche size and duration.
We interpret the results from a renormalization group perspective to point out that despite having broad tails, this is not related to a critical transition but rather similar to fluctuations driven by the reversal of the magnetic field in a ferromagnetic system. Another example of persistent correlation of fluctuations of a non critical system is constructed, where a particle undergoes Brownian motion on a slowly varying potential. 
\end{abstract}

\maketitle

\section{Introduction}

The idea that the brain operates at a critical point of a phase transition has been present in the field of Neuroscience since the first experimental description of neuronal avalanches, with its size and duration scale-free bursts of neuronal activity, by Beggs and Plenz \cite{Beggs11167}. Further extensive activity has discussed and given theoretical support   \cite{Kinouchi2006,schuster2014criticality,Girardi-Schappo_2021,subsampled,RevModPhys.90.031001,Mauro2,DelPapa} to the \emph{critical brain hypothesis}. A possible evolutionary reason for criticality is that it has been shown to be related to optimal information processing, in the sense of maximum dynamic range in neural networks of excitable units that model sensory systems, as shown by Kinouchi and Copelli \cite{Kinouchi2006}, where criticality is associated to the network topology, i.e. bond percolation of the couplings. Self organized criticality can occur due to learning in a recurrent neural network, driven by random inputs \cite{DelPapa}.  However,  experimental or numerical signatures such as power laws and scaling are possible without criticality, with examples \cite{GISIGER,Bedard,Touboul} in general areas of biology or in particular in neural networks models. 
It is interesting to further investigate putative critical behavior in systems with slowly decaying correlations and broad tail distribution.

\begin{figure}
     \centering
     \begin{subfigure}[b]{0.45\textwidth}
         \centering
          \includegraphics[width=\textwidth]{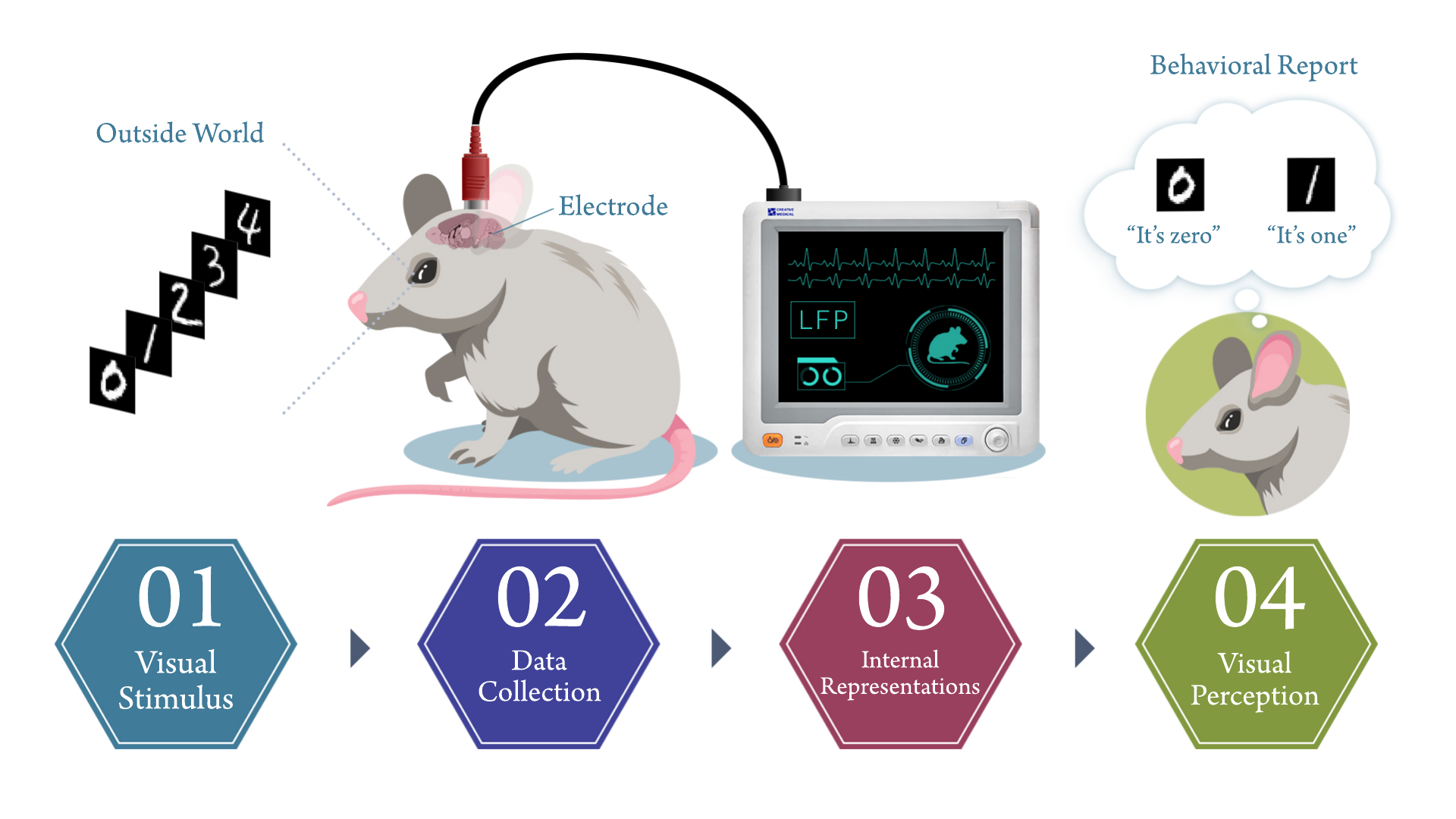}
          \caption{}
     \end{subfigure}
     \hfill
     \begin{subfigure}[b]{0.4\textwidth}
         \centering
         \includegraphics[width=\textwidth]{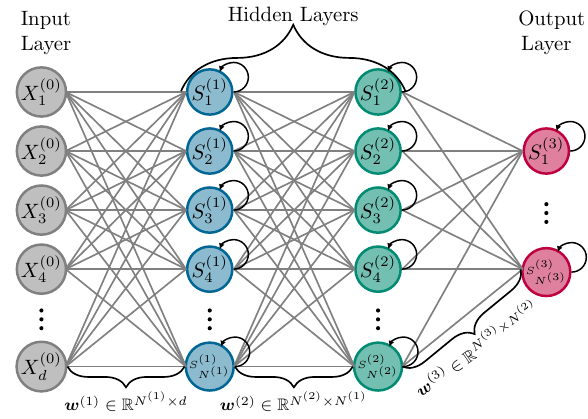}
         \caption{}
     \end{subfigure}
     \caption{(a) Typical experimental setup, using electrodes for studying cortical activity during a visual perception task. (b) This experiment is simulated with an artificial neural network setup. The diagram shows the fully connected feed forward spiking neural network used in this paper. Each unit is a LIF spiking neuron and the loop arrow symbolizes the time-dependent changes of the membrane potential. The "electrodes" collect information from the hidden layers.}
    \label{fig:experimental_setup}
\end{figure}

In this paper we investigate the statistics of fluctuations in the internal layers of a feed-forward Neural Network of spiking neurons when the network is performing a ``meaningful cognitive'' task of pattern recognition. This is done (see details in section \ref{material}) to mimic the data collection in an experiment of cortical implanted electrodes in rats exposed to a free environment illustrated in Figure \ref{fig:experimental_setup}.
We study the time truncated-correlation of fluctuations of the average spike activity, of trains of spikes and average membrane potentials of the internal representations in the network's hidden layers and measure their characteristic decay times, generically denoted $\xi$. Large values, as are typically associated to criticality,  can be seen when the  time dependent input changes and the classifier network changes the classification.  Networks exposed to a sequence of different images in the same category do not present such "diverging" (much  longer) $\xi$'s. We also investigate size and duration of neuronal activity avalanches. These broad-tailed distributions are apparently similar to those  which have been summoned to justify the term critical, are only found when the network classifies a changing environment. However, despite the allure of calling this a (near-) critical state, we are doubtful about this interpretation. In section \ref{broadtails} we show a simple dynamical model with broad tailed distribution of fluctuations of its state,  associated to the shifts of the basin of attractions, similar to the dynamics the network undergoes, when the category of the input changes. This adds to the models that shows that long tailed distributions can occur without critical behavior. This very low dimensional system serves only as a simple metaphor for the appearance of broad tails, so we search for systems with non critical collective behavior with broad tails and their description using Statistical Mechanics and the Renormalization Group.

The methods of Statistical Mechanics have been extensively applied \cite{hopfield1982neural,Amit,Gardner1988,Engel01,Biehlshallow} to study computational properties of artificial neural networks. This allows the description of brain inspired dynamical systems in the language of emergent collective phenomena. The Statistical Mechanics description of emergent properties of the  thermodynamics state is analogous to concept formation \cite{Hofstadter, hopfield1982neural, Amit}. 
The Renormalization Group (RG) is the main theoretical tool to analyze systematically the multiple scales, spatial and/or temporal, relevant in  the critical region of a second order phase transition. In general the RG is a map from measures to measures or between Hamiltonians. Here we adopt the simplifying description as a map between configurations, which despite being limited in the scope of what the RG is, avoids discussing peculiarities of the renormalized measures or Hamiltonians, which may only occur, if at all, in the thermodynamic limit.  As no single scale is dominant, power laws are associated to the critical state. Since there are simpler theoretical methods useful in the noncritical regions of parameter space, there is a tendency to associate the RG to the study of the critical region, despite being useful elsewhere. Globally we can look at the RG as a classifier which maps microstates into order parameters that characterize thermodynamics states in the whole range of parameters. 
The evolution of probability distributions under exact RG transformations \cite{Wegner, PedroAriel} are an example of entropic dynamics \cite{Ariel2015}. The RG works by systematically marginalizing the degrees of freedom of a joint distribution of the degrees of freedom in all scales. The distribution of degrees of freedom at the coarser scale results from a MaxEnt implementation of constraints imposed by the coarsening procedure. This amounts to implementing a filtering process, where the filters are chosen {\it a priori}. A feed-forward neural network, with either deep or shallow architecture, also implements a filtering process \cite{MehtaSchwab,KochJanusz2018, Liwang,neuralMCRG_ChungKao, NestorEdnna}, with the difference that the filters are designed automatically during the learning process and not imposed by prior knowledge. Of course, for the RG,  prior knowledge about the adequate filters came from the seminal work of Kadanoff  \cite{Kadanoff}, Wilson \cite{WilsonKogut} and others. This identification is specially clear in the Monte Carlo version of real space RG (MCRG) \cite{SKMA, Swendsen} which we discuss in \ref{secMCRG} as an example of a convolutional neural network. An application of the MCRG to a 2d Ising model with a slowly changing external field, shows that large susceptibility, and the broad tail distribution of fluctuations it entails, can occur despite being clearly away from the critical region. In the case where in addition to the external field, there is a random field contribution at each site (normal,$\mu=0, \sigma= R$) and $T=0$ \cite{SethnaBarkhausen}, \cite{Sethna2001}, the field reversal dynamics  is only truly critical with power laws of the Barkhausen noise at a critical value $R$.
The transition is due to the reversal of the external magnetic field, which plays the role of the input pattern to the neural network.  In our study, broad-tailed distributions  are not associated to a critical transition, but rather to the shifting nature of basins of attraction that accompanies changes in the images presented  to the network. 

\section{Materials and Methods \label{material}}

\subsection{Experimental models: the neural networks}
We constructed a classifier  using a {\it deep} architecture, i.e a multilayered feed-forward fully connected, from layer to layer, neural network of spiking neurons shown in Figure \ref{fig:experimental_setup}.b.
It was subject to supervised learning with the goal of recognizing images from standard datasets, the MNIST and the Fashion-MNIST. Both tasks had the same structure with respect to data set size, training and testing splits, composed of 60.000 training and 10.000 testing 28x28 gray scale images of 0 to 9 handwritten digits and of 10 categories of fashion products, respectively.

We use a simple and general purpose neuron model for large-scale training of artificial neural networks. The leaky integrate-and-fire (LIF) neuron, that goes back to Lapique \cite{Lapique} (see \cite{Abbott1999LapicquesIO,Dayan}), captures the essential behavior of the electric potential difference in nerve cells, it is computational cheap and adequate for our purposes. The LIF  follows the principle of integration of inputs combined with a reset mechanism. The dynamics of each $i^{\text{th}}$ LIF neuron in the $l^{\text{th}}$ layer, with a reset by subtraction of threshold $\theta$, is written as a discrete time step equation

\begin{equation}\label{eq:membrane_potential}
\begin{dcases}
    S^{(l)}_{i}[t] =  \Theta\left(U^{(l)}_{i}[t] - \theta\right) = 
    \begin{cases}
        1 &\text{if $U^{(l)}_{i}[t] > \theta$}\\
        0 &\text{otherwise,} 
    \end{cases}
    \\
    U^{(l)}_{i}[t+1] = \beta U^{(l)}_{i}[t] + \sum_{j=1}^{N^{(l-1)}}w^{(l)}_{i j}S^{(l-1)}_{j}[t+1] - \theta S^{(l)}_{i}[t],
\end{dcases}
\end{equation}
where $S^{(l-1)}_{j}$ is the presynaptic input of the $j^{\text{th}}$ neuron in the previous layer $(l-1)$, $w^{(l)}_{i j}$ is the adjustable synaptic coupling between the $j^{\text{th}}$ neuron in the previous layer $(l-1)$ and the $i^{\text{th}}$ neuron in layer $(l)$ and $\beta$ controls the membrane potential decay.

\begin{figure}
    \centering
    \includegraphics[width=\textwidth]{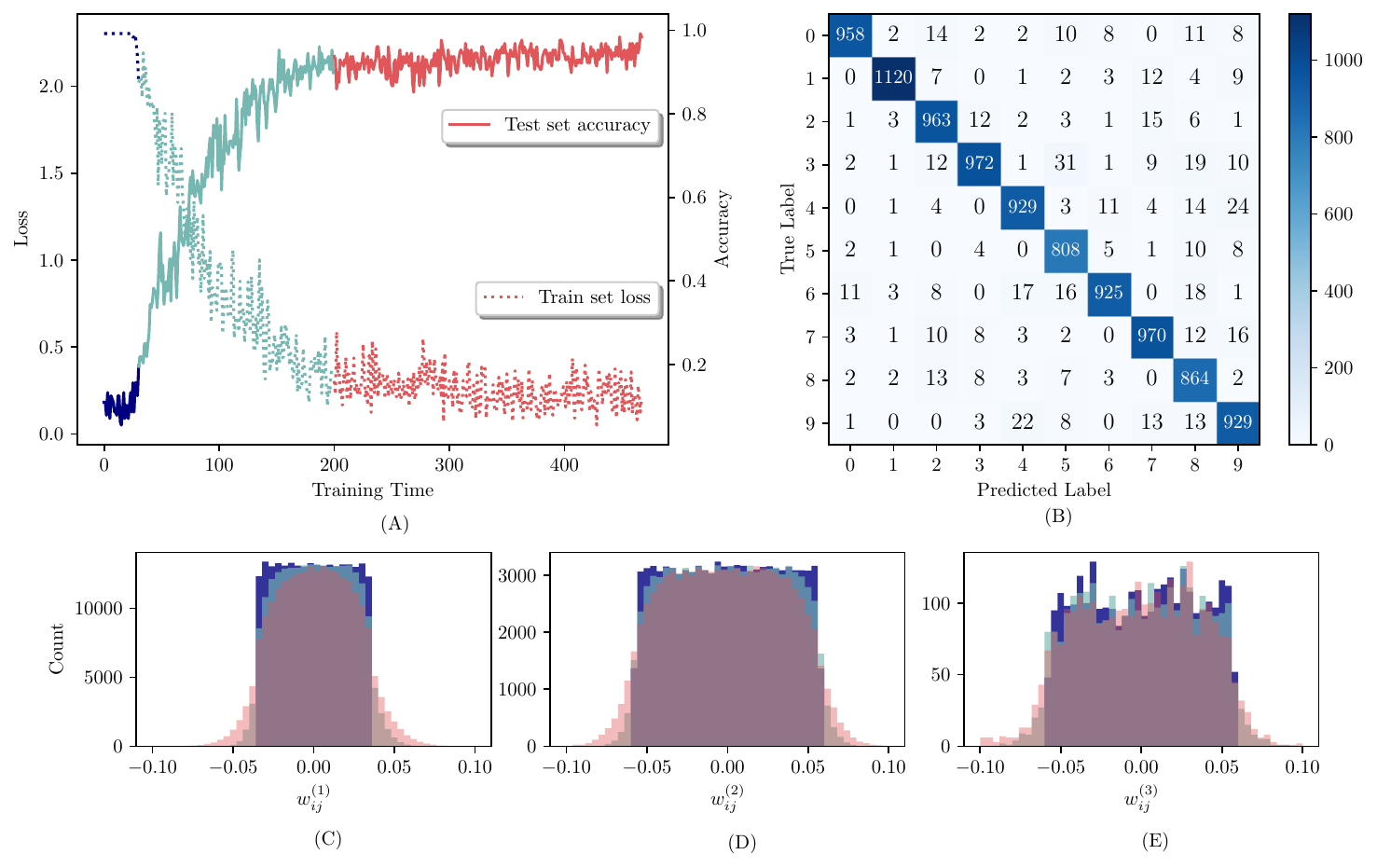}
    \caption{Training session over MNIST train and test set. (A) Loss (left axis) and Accuracy (right axis). Three different phases can be seen: an initial plateau or slowly rising accuracy phase, a fast rising and a saturated plateau. (B) Confusion matrix over the test set after the training. (C), (D) and (E) The evolution of the empirical distributions of weights between the layers,  during training, show no sign of topological criticality. The results for the Fashion-MNIST data set are not shown since they are not significantly different.
    \label{fig:fit_mnist}}
\end{figure}

\begin{figure}[H]
    \centering
    \includegraphics[width=0.7\textwidth]{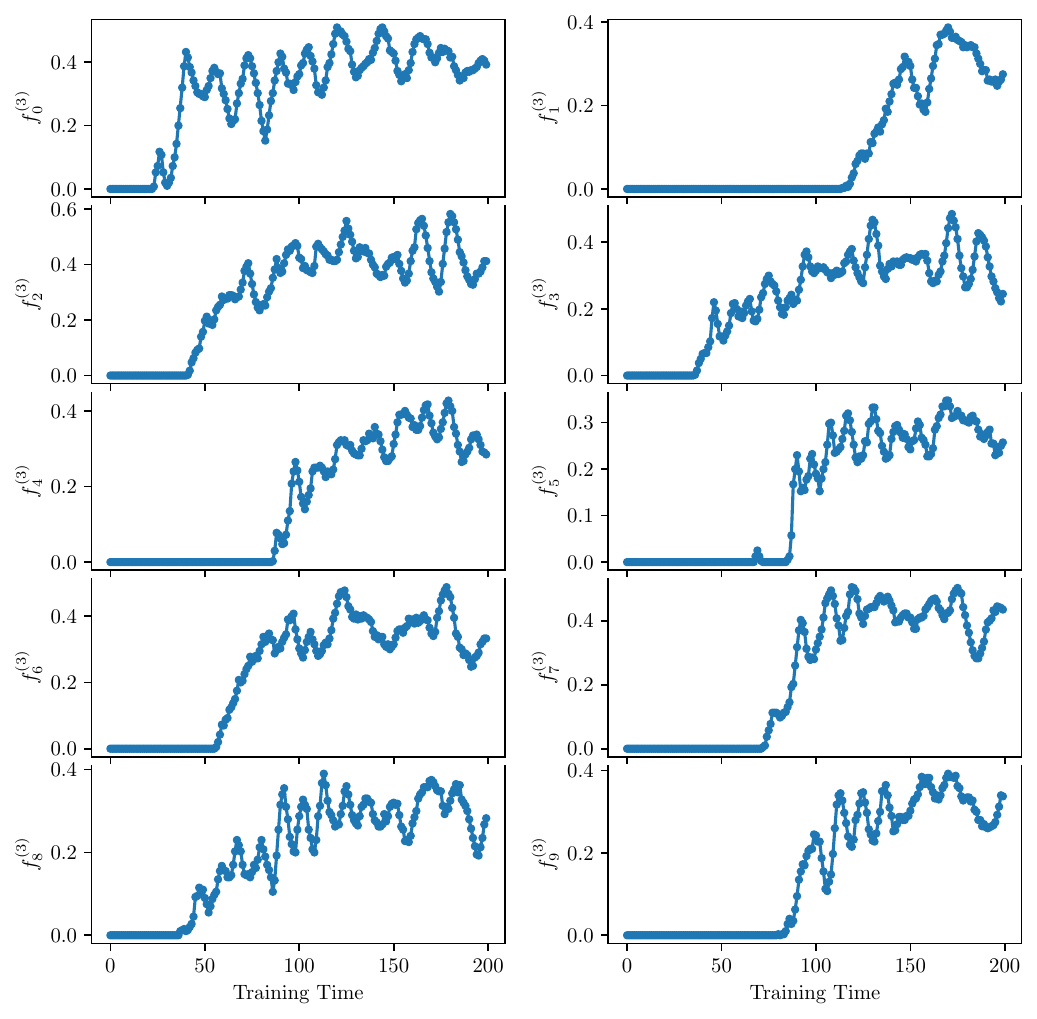}
    \caption{Evolution with training time of the firing rate $f_k^{(3)}$ of output neuron $k$ when input patterns of type $k$ are presented. Note that it takes different training times until the network can identify different categories. MNIST dataset.}
    \label{fig:recognition_training}
\end{figure}

To determine an appropriate learning algorithm for this class of neural network, neuronal communication must be precisely defined. Rate coding relates to the transmission of meaningful information only by means of neuron firing rates, i.e, the response of spike train frequency increases as the sum of weighted input increases. We implemented rate coding, which can be simply adapted to our model. The output layer has $n=10$ neurons, so that in the classification problem each individual response of output neurons is connected to the "recognition" of the category of the input image: `It is a zero' or `It is a shoe'. This is the behavioral report, a visual perception of a labeled input. For training, we used the error backpropagation through time (BPTT) \cite{werbos1988generalization} with the surrogate gradient approximation \cite{surrogate_gradient, snnTorch}. $S = \Theta(U(t) - \theta)$, as a function of $U$ is a step function, and as a function of $t$ is a spike, firing when the potential becomes larger than the threshold and coming back to zero after the subsequent reset of the potential. This function is substituted by a differentiable function $\tilde{S}$, to be used only in the backward computation step.  A component of the gradient of the loss function $\mathcal{L}$ is:
\begin{equation}\label{eq:surrogate_gradient}
     \frac{\partial \mathcal{L}}{\partial w_{ij}^{(l)}} = \sum_{t=0}^{t_{U}}\frac{\partial \mathcal{L}(t)}{\partial S_{j}^{(l)}(t)} \frac{\partial \tilde{S}_{j}^{(l)}(t)}{\partial U_{j}^{(l)}(t)}\frac{\partial U_{j}^{(l)}(t)}{\partial w_{ij}^{(l)}},
\end{equation}
where $t_{U}$ is an integration time over the dynamics of the membrane potential. For the gradient, we use a threshold-shifted sigmoid curve,  
\begin{equation}
    \tilde{S} = \frac{1}{1 + e^{\theta - U(t)}}, \hspace{1cm} \frac{\partial \tilde{S}}{\partial U} = (1-\tilde{S})\tilde{S}.
\end{equation}
\begin{table}
\begin{center}
    \begin{tabular}{c|c}
        \hline
        \hline
        \multicolumn{2}{c}{Hyperparameters}\\
        \cline{1-2}
        Neural Network    & Neuron\\
        \hline
        $d = 28\times 28$      & 1st Order Leaky Integrate and Fire Neuron \\
        $N^{(1)} = 300,$  $N^{(2)} = 300,$  $N^{(3)} = 10$      & $\beta = 0.5$ \\
        $\text{Batch size} = 128$ & $\theta = 1$ \\
        $\text{Epochs} = 1$ &$t_{U} = 25$  \\
        $\eta = 5 \times 10^{-4}$ & \\
        Optimizer: Adam &  Reset Mechanism: Subtract  \\
        Accuracy metric: Spike Count  & Backward pass: Fast Sigmoid  \\
        Loss: Cross Entropy Spike Count & \\
        \hline
        \hline
    \end{tabular}
    \caption{\label{tb:hyperparameters_training} Hyperparameters used on training. We used standard training for networks with $N^{(1)},  N^{(2)}$ neurons in the 2 hidden layers. }
\end{center}
\end{table}
Hyperparameters were chosen according to table \ref{tb:hyperparameters_training}. The training section was composed by an epoch, a pass over the training set, where the neural network attained an accuracy above of $80\%$ depending of dataset. Accuracy is measured by attribution the classification label to the output neuron with the highest firing rate. Figure \ref{fig:fit_mnist} shows the loss and accuracy as a function of the number of subsets (mini-batches) of previously randomly shuffled training set and test set. Training time refers to each interaction over mini-batches of the dataset. The confusion matrix, a table of true positive and false positive matches according to the data labels is shown in Figure \ref{fig:fit_mnist}.C.
 Training shows three distinct phases. A short initial phase where the network still cannot implement the task, a fast rising effective learning phase, and an almost saturated phase showing good performance. After training, the confusion table becomes practically a diagonal matrix, indicating that the model classification is most likely a true positive. Figure \ref{fig:recognition_training} shows the firing rates of the correct output showing different learning time thresholds for the different categories being learned.

\subsection{The mathematical characterization}
We use the following notation for the time average at time $t$ of any quantity $f$ in a time window of width $\Delta t$
\begin{equation}
    \Bigl \langle \, f(t) \, \Bigr \rangle = \frac{1}{\Delta t}\sum_{t'=t}^{t + \Delta t} \, f(t') \,.
\end{equation}
When $f$, for fixed $t$, is an array, so is the above expression and the sums are point wise. The fluctuation of a quantity is $\overline{f} =f - \langle f \rangle$. The un-normalized truncated correlation is   
\begin{equation}
    c_{f,g}(t,\tau,\Delta t) =\Bigl \langle \, \overline{f(t)}\,\,\overline{g(t-\tau)} \, \Bigr \rangle,
\end{equation}
 $\overline{f}\,\, \overline{g}$ is a dot product for arrays, 
 and the normalized correlation of fluctuations or the truncated correlation  is  
 \begin{equation}
     C_{f,g}(t,\tau,\Delta t)= \frac{c_{f,g}}{\sqrt{c_{f,f}c_{g,g}}}. \label{correlacoes}
 \end{equation}
The spike activity in the $l^{\text{th}}$ layer is described by an $N^{(l)}$-dimensional array at time $t$
\begin{equation}
    \bm{S}^{(l)}(t) = \left[S_{1}^{(l)}(t), \cdots, S_{N^{(l)}}^{(l)}(t)\right],
\end{equation}
where $N^{(l)}$ is the number of neurons and $S_{i}^{(l)} = \{0,1\}$, the binary output specifying whether neuron $i$ has fired at $t$.
To  characterize the fluctuations of internal representations in layers $l^{\text{th}}$ and $k^{\text{th}}$ we use $C_{\bm{S}^{(l)},\bm{S}^{(k)}}(t,\tau,\Delta t)$.

We also define the following  macroscopic variables (aggregated): the average activity of the $l^{\text{th}}$ layer at time $t$  
\begin{equation}
    \rho^{(l)}(t) = \frac{1}{N^{(l)}}\sum_{i=1}^{N^{(l)}}S^{(l)}_{i}(t),\label{rho}
\end{equation}
and the average membrane potential in the $l^{\text{th}}$layer at time $t$
\begin{equation}
    u^{(l)}(t) = \frac{1}{N^{(l)}}\sum_{i=1}^{N^{(l)}}U^{(l)}_{i}(t).
\end{equation}
In the simulations we measure $C_{\bm{S}^{(l)} \bm{S}^{(l)}}(t; \tau, \Delta t) , C_{\rho ^{(l)} \rho^{(l')}}(t; \tau, \Delta t)$ and $ C_{u^{(l)} u^{(l')}}(t; \tau, \Delta t)$, functions of the three parameters $t$, $\tau$ and $\Delta t$. From an analogy to spins systems, we expect these correlations to decay with $\tau$ as the product of a fast decaying exponential with characteristic time $\xi_\tau$ and a slow decaying algebraic function. If criticality is approached $\xi_\tau$ grows and there is a crossover to purely algebraic decay, but only in the thermodynamic limit. For finite systems an exponential function is a good model and we consider
\begin{equation}\label{eq:exponential_decay}
    C(\tau; t, \Delta t) \sim \exp{\Biggl \{-\frac{\tau}{{\xi_{\tau}}}\Biggl\}}.
\end{equation}
$\xi_{\tau}$ plays a similar role as the correlation length.  Fixing $t$ and $\Delta t$, we can see how the temporal correlation varies with $\tau$. The correlation time can be computed for each time step $t$ from a fit to equation \ref{eq:exponential_decay}. 

We also define what we mean in this context by avalanches of activity and analyze their distribution.
We say that a $\rho$ avalanche in the hidden layers, is happening when $\rho(t)= \sum_l\rho^{(l)}(t)$, summed over hidden layers, is above its time average $\bar R$. An avalanche's size is the integrated area below $\rho(t)$ between the crossings to $\rho> {\bar R}$ and back to  $\rho < {\bar R}$, and the duration is the difference in time between those consecutive occurrences. 
\subsection{Simulations \label{simulationsNN}}
 We use "movies", sequences of images described in Table \ref{TableMovie} as inputs to the NN during the simulations.
The set of numerical experiments  is shown in Table  \ref{Tablesimulations}. $\alpha$ is a parameter introduced to simulate partial blocking of the input to the network. Every pixel of an input pixel is multiplied by $\alpha$ in the interval $(0,1]$. It is a simplistic model of the effect of anesthesia acting only on the input layer. It is used just to study the pattern recognition of a dim input and not as a model of the effects of a particular type of anesthesia itself. 
\begin{table}
\caption{The "movies" \label{movies}}
\begin{tabularx}{\textwidth}{ccccc}
\toprule
{Movie type}& {Composition}	&Description & \textbf{$\delta t$} & Measures\\
\midrule
$M_0$& Random inputs		& New random image every time step		& none & $\xi_\tau$ \\
\hline
$M_S$ &2 or more classes 		& Few slow transitions between classes&  $\delta t \approx 400-500$&$\xi_\tau$\\ \hline
$M_F$& 2 or more classes  & Several fast transitions between classes & $\delta t\sim t_U \sim 25$ & Avalanches \\
\bottomrule
\end{tabularx}
\caption*{For $M_S$ and $M_F$, $\delta t$ is the number of presentations of random choices of examples in the same class between transitions. It measures the variability of the environment presented to the NN.\label{TableMovie}}
\end{table}

\begin{table}
\caption{Simulations.\label{Tablesimulations}}
\begin{tabularx}{\textwidth}{ccc}
\toprule
\textbf{Training Phase}	& \textbf{Movie}	& \textbf{Conditions}\\
\midrule
early 		& $M_0$		&  $\alpha =1$\\
intermediate& $M_0, M_S, M_F$& $0< \alpha \leq 1 $\\
saturated & $M_0, M_S, M_F$ & $0< \alpha \leq 1 $\\
\bottomrule
\end{tabularx}\caption*{ The intensity of the  "anesthesia" is $1-\alpha$.  }
\end{table}

\section{Results }
\begin{figure}
     \centering
     \begin{subfigure}{.40\textwidth}
    \centering{\label{fig:temporal_corr_tau_frames_mnist}\includegraphics[width=\linewidth]{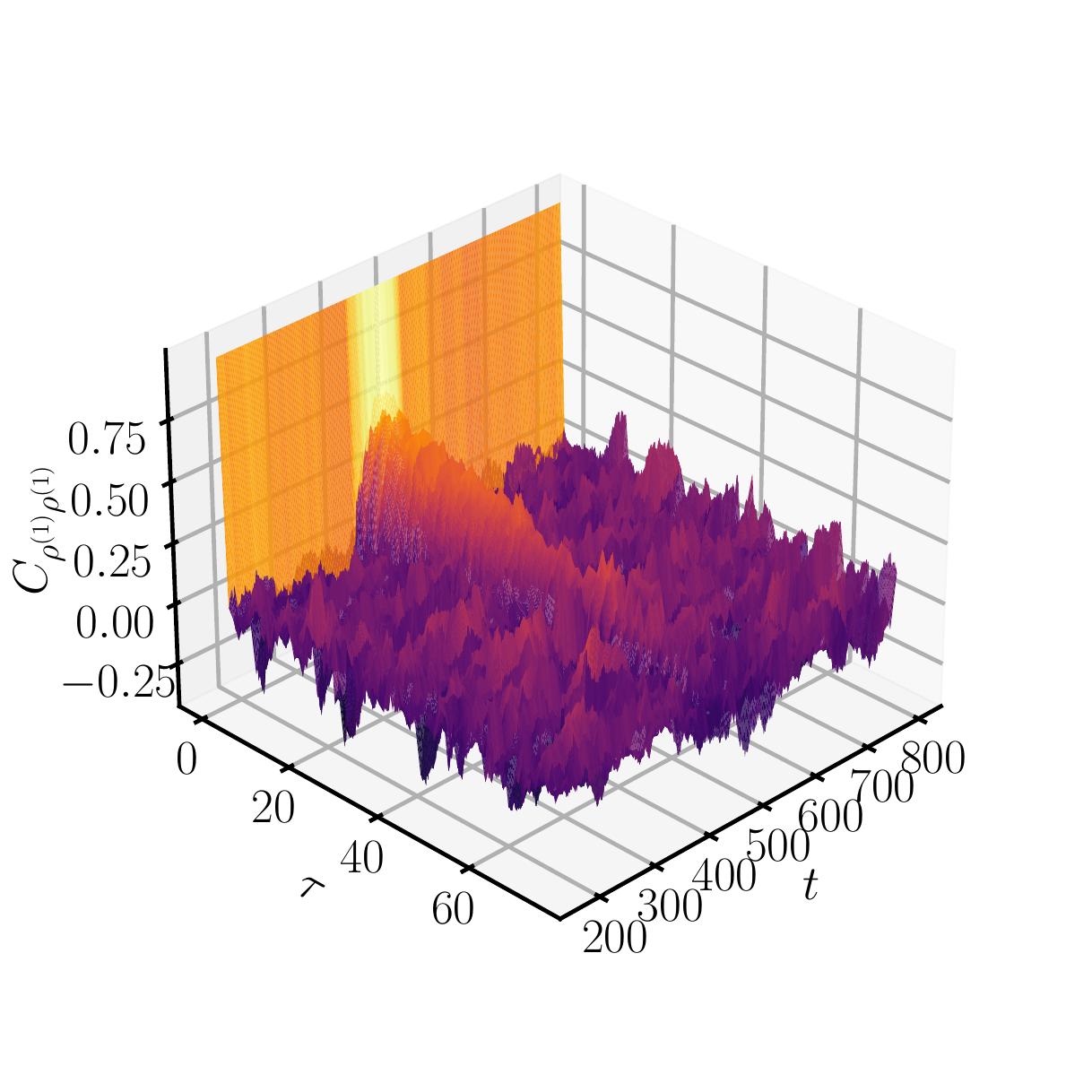}}\hfill
    \end{subfigure}
     \begin{subfigure}{.40\textwidth}\centering{\label{fig:temporal_corr_tau_frames_fit_mnist}\includegraphics[width=\linewidth]{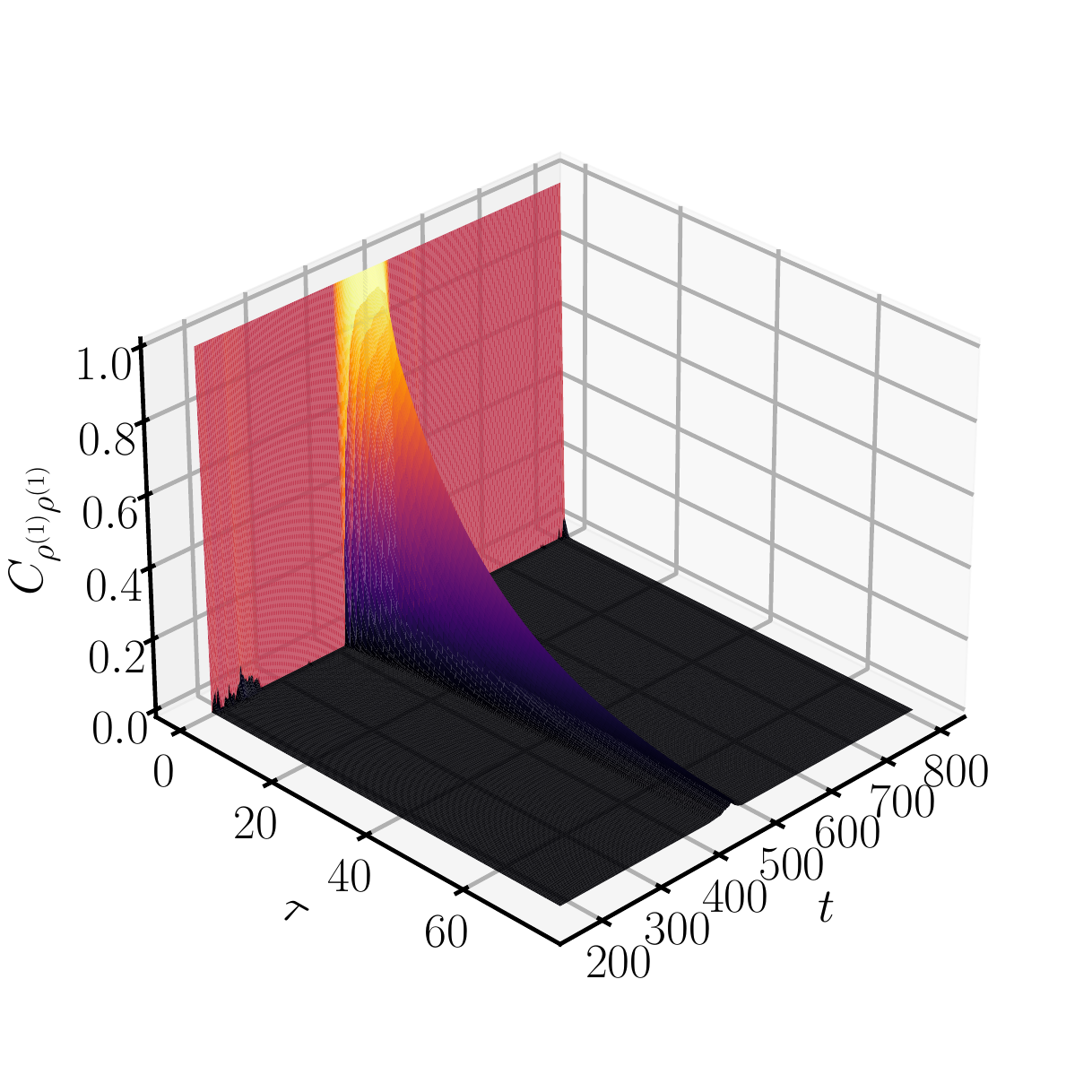}}
   \end{subfigure}
   \caption{IR fluctuations temporal correlation as a function of $\tau$ measured on a sliding window centered at time $t$ for the single image change  movie. Left: the raw measured correlations. Right: For each $t$, the best fit of an exponential decay $e^{-\frac{\tau}{\xi_\tau}}$, equation\ref{eq:exponential_decay}. At the region of the transition of images, the decay time $\xi_\tau$ increases a few orders of magnitude  \label{fig:temporal_corr_tau_frames_mnist_all}}
   
\end{figure}

\begin{figure}
    \centering
    \includegraphics[width=\textwidth]{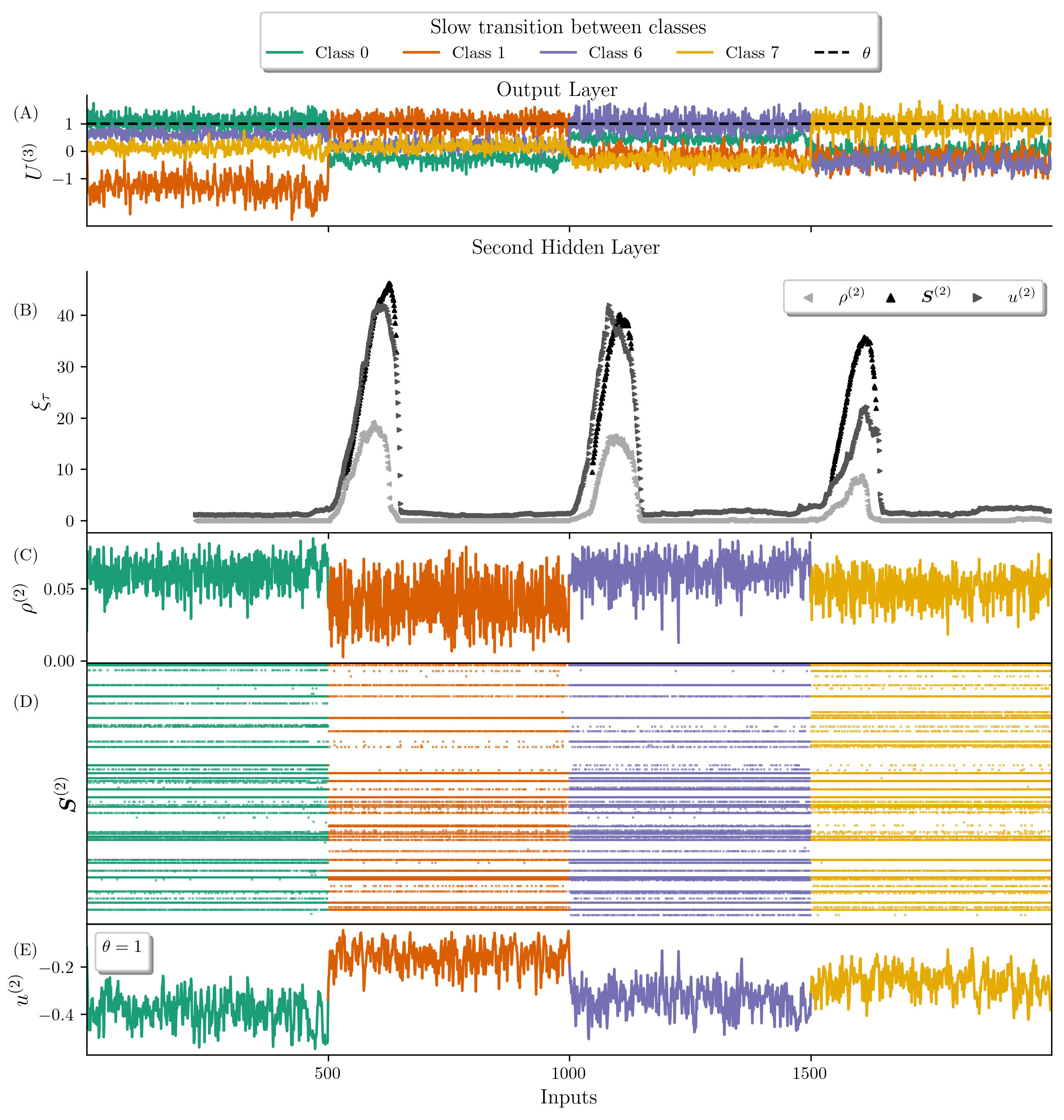}
    \caption{The $M_S$ movie input (no anesthesia $\alpha=1$) contains a sequence of 500 random images of label "0" (green), then 500 of label "1" (orange), 500 of label "6" (purple), and 500 of "7" (yellow) from the MNIST database.  (A) The membrane potential of output neurons responsible for classification of labels during the movie, (B) The characteristic time of decay of correlations of the fluctuations of the average activity, of spikes and average membrane potentials of the second hidden layer. (C) Average activity. (D) Spike train: raster plot of a few neurons of hidden layer 2. (E) Average membrane potential.}
    \label{fig:slow_movie}
\end{figure}

At a given point in the training time (Figure \ref{fig:fit_mnist}) we study the IR that occur when a "movie"  is presented to the NN. We start analyzing the fully trained network. Several types of movies can be presented, see Table \ref{TableMovie}. Movie $M_0$ presents no interest at all since random images are shown to the network, which correctly fails to identify any category. These results, which only serves as a control test, are not shown. Next we present a movie of $M_S$ type. For intervals comprising  $\sim 500$ frames random examples of one of the ten categories is shown, then for another $\sim 500$ frames, examples from another category are presented. Nothing interesting happens except for a small region in time starting at the transition with large and persistent fluctuations, see Figure \ref{fig:temporal_corr_tau_frames_mnist_all}. 
In Figure \ref{fig:slow_movie} we show results from a typical run with several input class changes. The NN is working satisfactorily as shown in \ref{fig:slow_movie}.A, since the output cell with the dominant $U^{(3)}$ is the correct one.  

 Figure \ref{fig:slow_movie}.B shows results for the second hidden layer IR. A very large increase in $\xi_\tau$ following the input label transitions. 
 A similar behavior, with smaller peaks, is seen for the IR in the first hidden layer.
 While the NN moves rapidly toward the new classification, IR show persistent fluctuations, around the transitions. Is this peak in time correlation length the signature of criticality? We argue in section \ref{broadtails} that it is not.

\begin{figure}
    \centering
    \includegraphics[width=\textwidth]{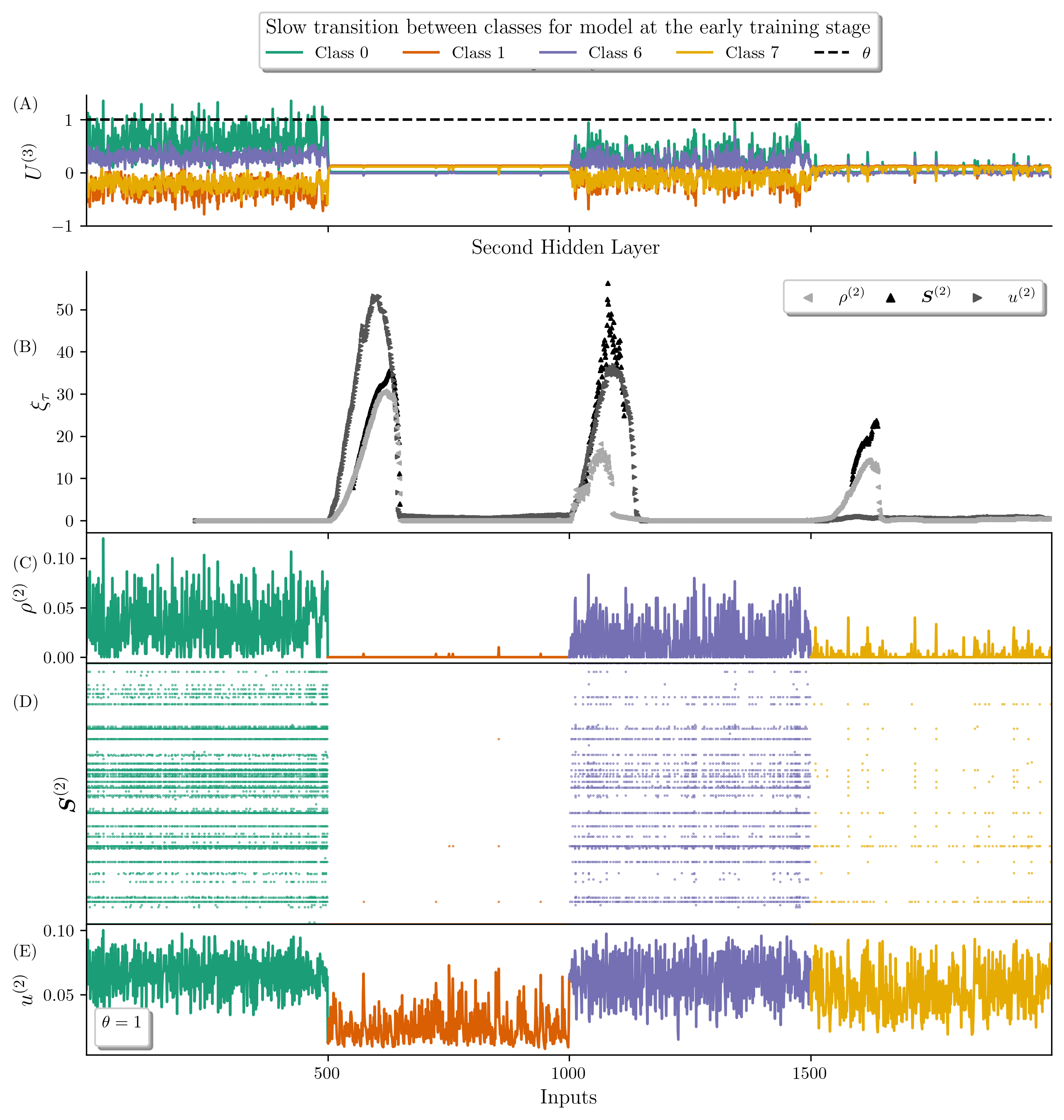}
    \caption{Same as Figure \ref{fig:slow_movie} but for a partially trained NN, where only "0" is recognizable. Transition regions show persistent fluctuations, independently of output recognition.}
    \label{figuraSemTreino}
\end{figure}

 A second numerical experiment investigates the results of partial training. We show results in Figure \ref{figuraSemTreino} for a training time of 25, when as Figure \ref{fig:recognition_training} shows, only category "0" is partially recognizable in the output layer. It may be expected that since the category attractors are not yet fully developed, persistent fluctuations should not appear. However, despite not achieving  correct classification over all data labels, such persistence occurs  during the transition of input categories. This is observed in distinct cases, first  during the transition of a correctly classified input to one not yet learned ("0" $\rightarrow$ "1"). Second, when the neural network still has not learned both input labels ("1" $\rightarrow$ "6" $\rightarrow$ "7"). This is evidence that during the learning process,  attractors of the dynamics are partially present in the hidden layer, while still sub-threshold in the output layer.
 
\begin{figure}
     \centering
     \includegraphics[width=\textwidth]{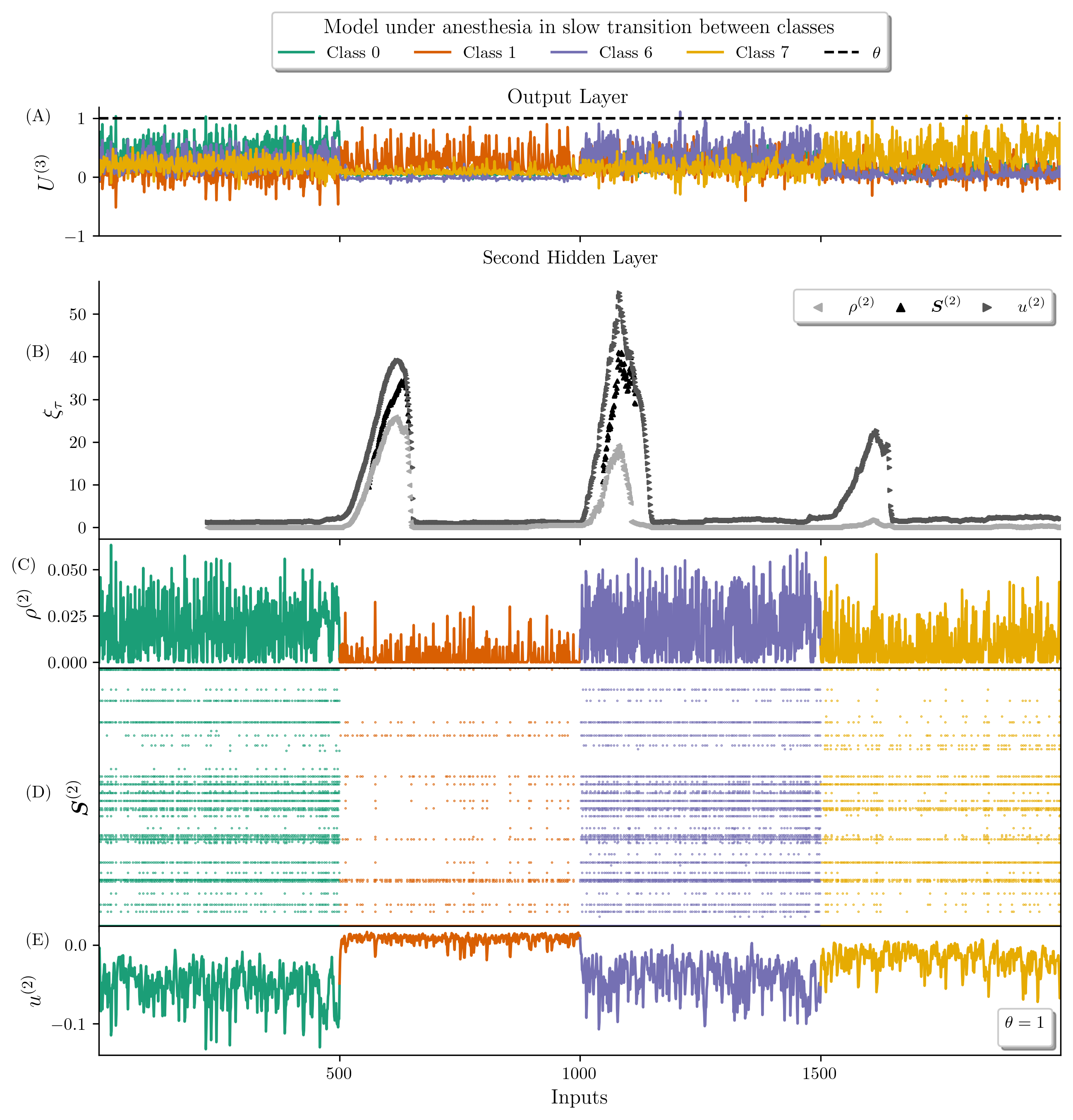}
     \caption{Same as Figure \ref{fig:slow_movie} but with neural network under effect of anesthesia, $\alpha=.4$, with a $M_S$ input. (A) The network is not able to correctly categorize the inputs, which generate sub threshold activity at the output layer: there is no communication of the computation to the outside world. 
     \label{anesthesia_slow_movie}. (B) Nevertheless, the transition of inputs generates persistent fluctuation correlation. (C, D) show the activity in the second hidden layer, but (E) $u^{(2)}$ fails to exceed the threshold. }
 \end{figure}

 In a  third experiment, the fully trained network, used in figure \ref{fig:slow_movie}, is presented to a $M_S$ movie with a dimmed input, every pixel input is multiplied by $\alpha<1$, which we call a model for anesthesia. Figure \ref{anesthesia_slow_movie} shows again that for  inputs on which the output is not correct due to diminished input sensibility, the sub-threshold images can elicit persistent correlations with large characteristic times.

\begin{figure}[H]
     \centering
     \includegraphics[width=\textwidth]{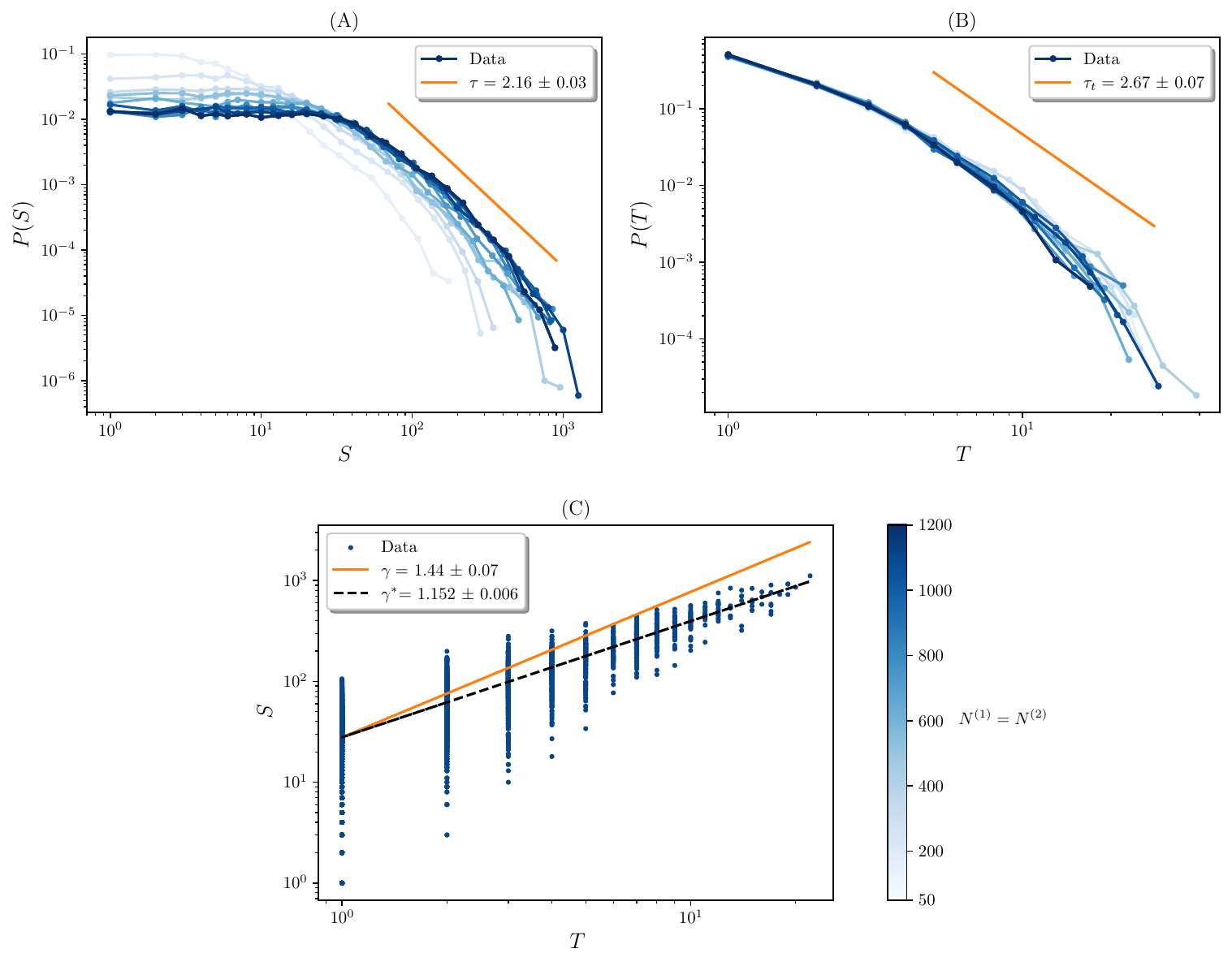}
     \caption{(A) Size and (B) duration avalanches. Power law fit using \cite{alstott2014powerlaw} are very sensitive to the choices of the minimum and maximum values.}
      \label{powerlaw}
     \label{fig:avalanches}
 \end{figure}

 We use the $M_F$ movie to generate several fast input transitions so that the final output alternates between different categories, to simulate a free moving agent in a rich environment. This input generates a large amount of variability in the IR and avalanches can be measured. Histograms are shown in Figure \ref{fig:avalanches}.A and B.  While we kept the number neurons in the hidden layers equal $N^{(l)}=N$ for all hidden layers, we ran simulations with different values of $N$ ($50$ and from $100, 200, \cdots$ to $1200$ .)
Each point in Figure \ref{fig:avalanches}.C shows the duration and size of an avalanche.  For the characterization of power laws we use a standard procedure \cite{alstott2014powerlaw}. The size and duration exponents $\tau, \tau_t$ give a  crackling exponent
 \cite{Sethna2001} 
 \begin{equation}
     \gamma = \frac{\tau_t-1}{\tau -1}\approx 1.44.
\end{equation}
 A direct fit of a power law in Figure \ref{fig:avalanches}.C yields a different  $\gamma^* \approx 1.15$. 
 The uncertainty of the fits, from \cite{alstott2014powerlaw} are too optimistic, being rather of the order of $.2$, not $0.03$. It is not possible to affirm that power laws are present in this case. Note that the procedures that yield these exponents are very sensitive, see e.g. \cite{Dallaportacopelli}, not only to choices of maximum and minimum values, but also to sub-sampling and to the noise process that is used to model the  fluctuations in the data. Since we are not defending this to be a signature of criticality we don´t delve into the details of the fit. If however the reader finds this a compelling argument for criticality, which we don't, it is clearly  not in the class of directed percolation.

\section{Broad tails without criticality \label{broadtails}}
\subsection{Langevin dynamics on a time dependent potential}
We now present a toy problem which is not critical in any sense and yet, it yields broad tail distribution in the style of what was found in the neural network of the previous section. 

\begin{figure}[H]
\includegraphics[width=10.5 cm]{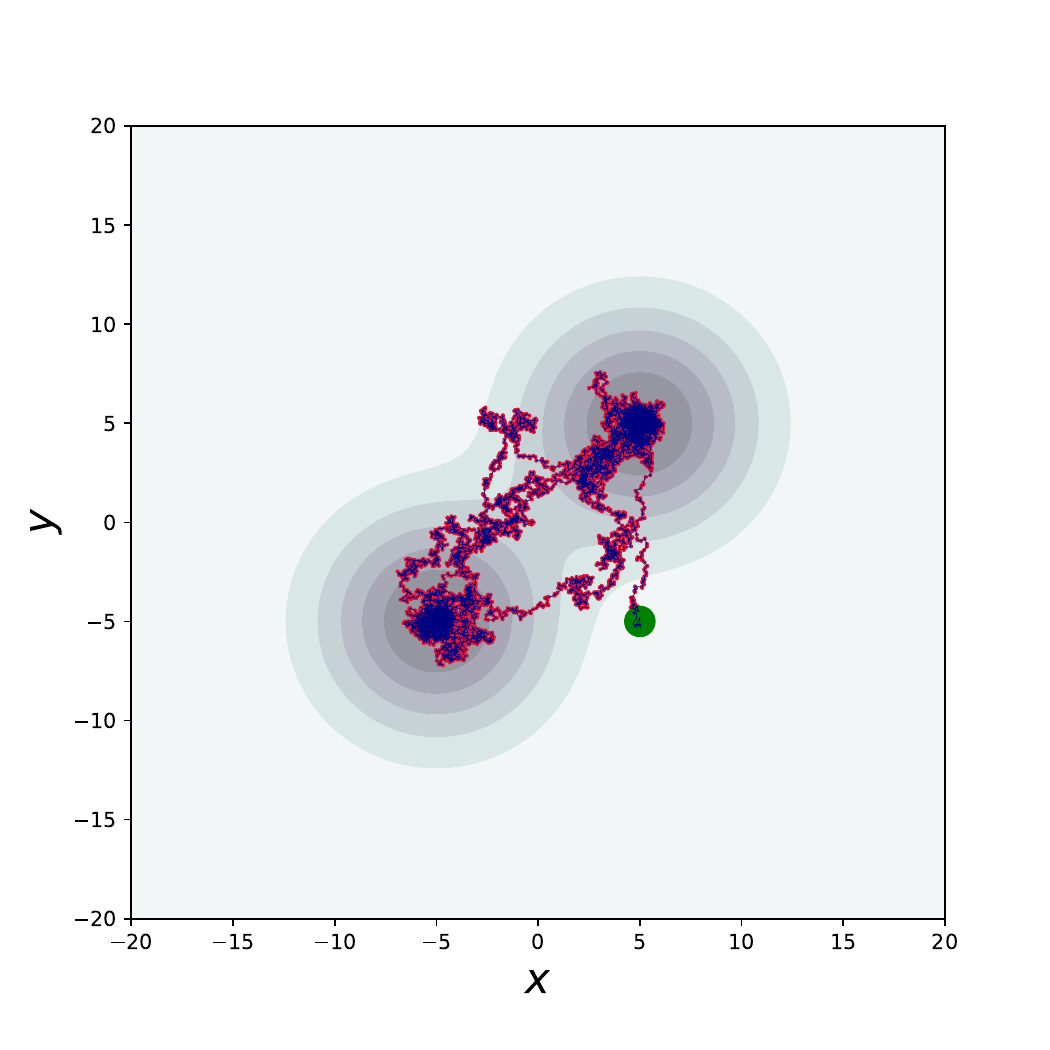}
\caption{\label{figEntre_pocos} A trajectory of the two dimensional model equation \ref{Langevin} starting from the green dot. It spends time in one well until it becomes very shallow and then wonders towards the other minimum that is becoming dominant.  The contours show the time average of the potential. Minima at $\pm{\bm r}_0= \pm(5,5)$, $\eta = 0.15, \sigma_W =0.1$.}
\end{figure}

\begin{figure}[H]
\includegraphics[width=.8\linewidth]{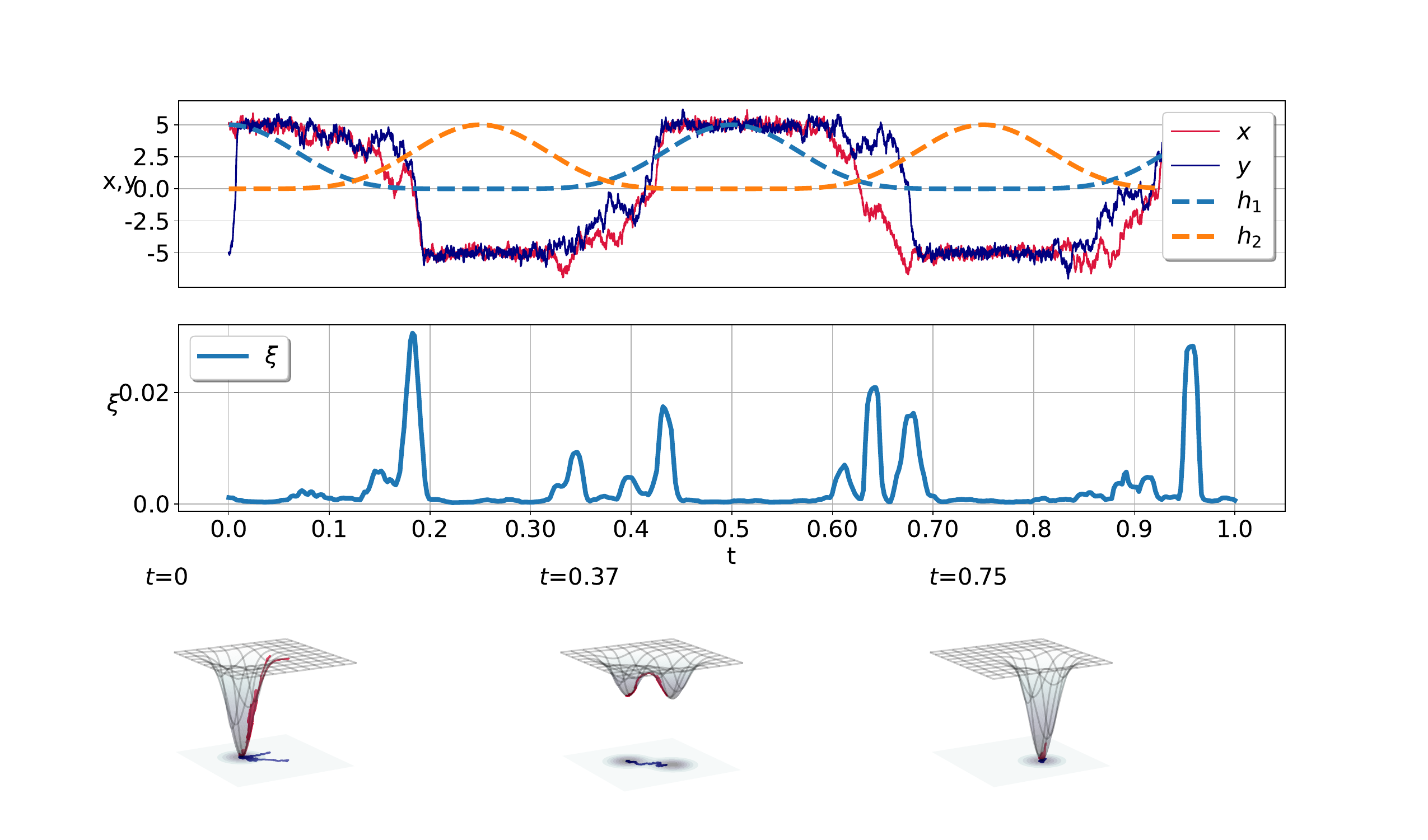}
\caption{Top: Coordinates $x$ and $y$ of a particle that evolves under the dynamics of equation \ref{Langevin}. $h_{1,2}$ are proportional to the depths of the wells. Center:  The continuous line shows a moving average of the  characteristic time $\xi$ of decay of the auto-correlation of the position of the particle, showing large peaks during the  transitions. Bottom: The 3d potentials drawn at the time  coordinate.\label{fig_3d} Moving average over window size $= 7$ ticks $=.5$ms)}
\end{figure}

With ${\bm r},{\bm r}_0 \in\mathbb{R}^2$, let $V({\bm r},\pm{\bm r}_0)$ be potential wells with a single minimum at $\pm{\bm r}_0$. We construct a time dependent potential 
\begin{equation}
    U(\bm r,t) = h_1(t)V({\bm r},{\bm r}_0)+h_2(t) V({\bm r},-{\bm r}_0),
\end{equation}
where for example we choose $h_1= |\cos(|\omega t)|)^k$ and $h_2= |\sin(|\omega t)|)^k$ and use as a typical example $k=6$. The potential changes from one well to another in periodic fashion, located at $\pm {\bm r}_0 $.
The process described by the discrete time Langevin dynamics
\begin{equation}
    {\bm r}_{t+\Delta t} = {\bm r}_t -\eta \nabla U({\bm r},t) +\bm W, \label{Langevin}
\end{equation}
with $W$ a two dimensional normal random process. $\bm r$ has a similar behavior to that of the IR of the previous section. It evolves from being around a minimum, drifting to the other as it becomes prominent, Figure \ref{figEntre_pocos}.
The dynamics obviously is not critical, nevertheless it has  long time persistence of the correlation decay when there is no clear minimum present, as shown in Figure \ref{fig_3d}. 

\unskip

\begin{figure}[H]
\includegraphics[width=.9\textwidth]{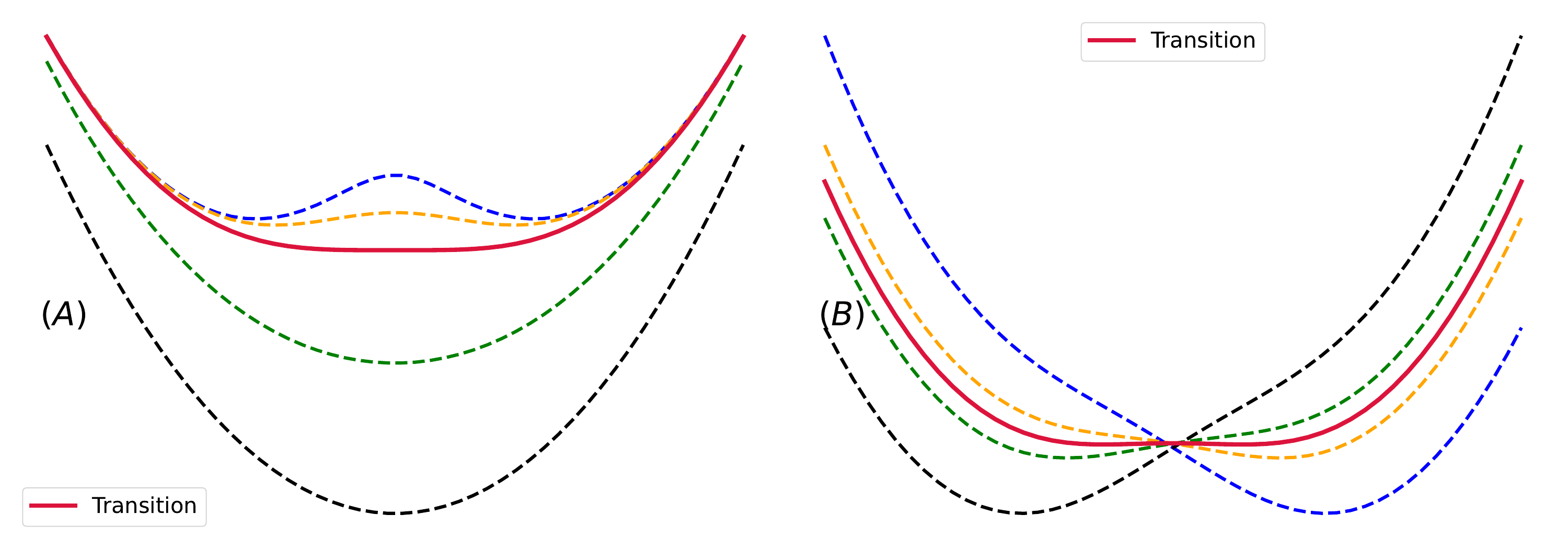}
\caption{(A) Free energy functional of an infinite range Ising model at zero external field, undergoing a critical second order transition between the disordered phase (black, green dashed curves) and the broken symmetry phase (orange, blue dashed lines). (B) Same model at temperature above critical. As the external field changes, the expected value of the magnetization transitions between two different ordered states ( $h>0$ orange and blue dashed lines, $h<0$ black and green dashed lines). For both graphs,  transitions points represented by the continuous red line.    \label{figTransicao}}
\end{figure} 
A second  analogy is based on the  infinite range Ising spin system in an external field, which plays the role of the input pattern to the neural network of section \ref{simulationsNN},
\begin{equation}
    H= - \frac{1}{2}(\sum_{1=1}^N  s_i)^2 + \sum_i h_is_i  
\end{equation}
At each site $i$ the magnetic field changes very slowly between the value $\xi_i^1$ and $\xi_i^2$. If the change is sufficiently slow, we can solve for the thermodynamics of the model for an external field fixed in time, and at each site the magnetization is obtained from 
\begin{equation}
    m_i(t) = \tanh(\beta(m_i(t)+h_i(t))\label{meanfield}
\end{equation}
If $h_i =0$, the exact solution is a minimum of the free-energy functional shown in Figure \ref{figTransicao}.A, with the typical second order transition occurring at the crtical value $\beta=1$. 
Now we consider a case where $\beta<1$, so the system is not critical and $h_i$ changes slowly in time. Then $m_i$ follows the minimum  of the curve shown in \ref{figTransicao}.B. In both cases, fluctuations measured in a time scale faster than the slow change of the external field, show broad tails independent of being critical or not.
\subsection{Monte Carlo Renormalization Group: convolutional classifier of the thermodynamic state \label{secMCRG}}
\begin{figure}[H]
\includegraphics[width=\textwidth ]{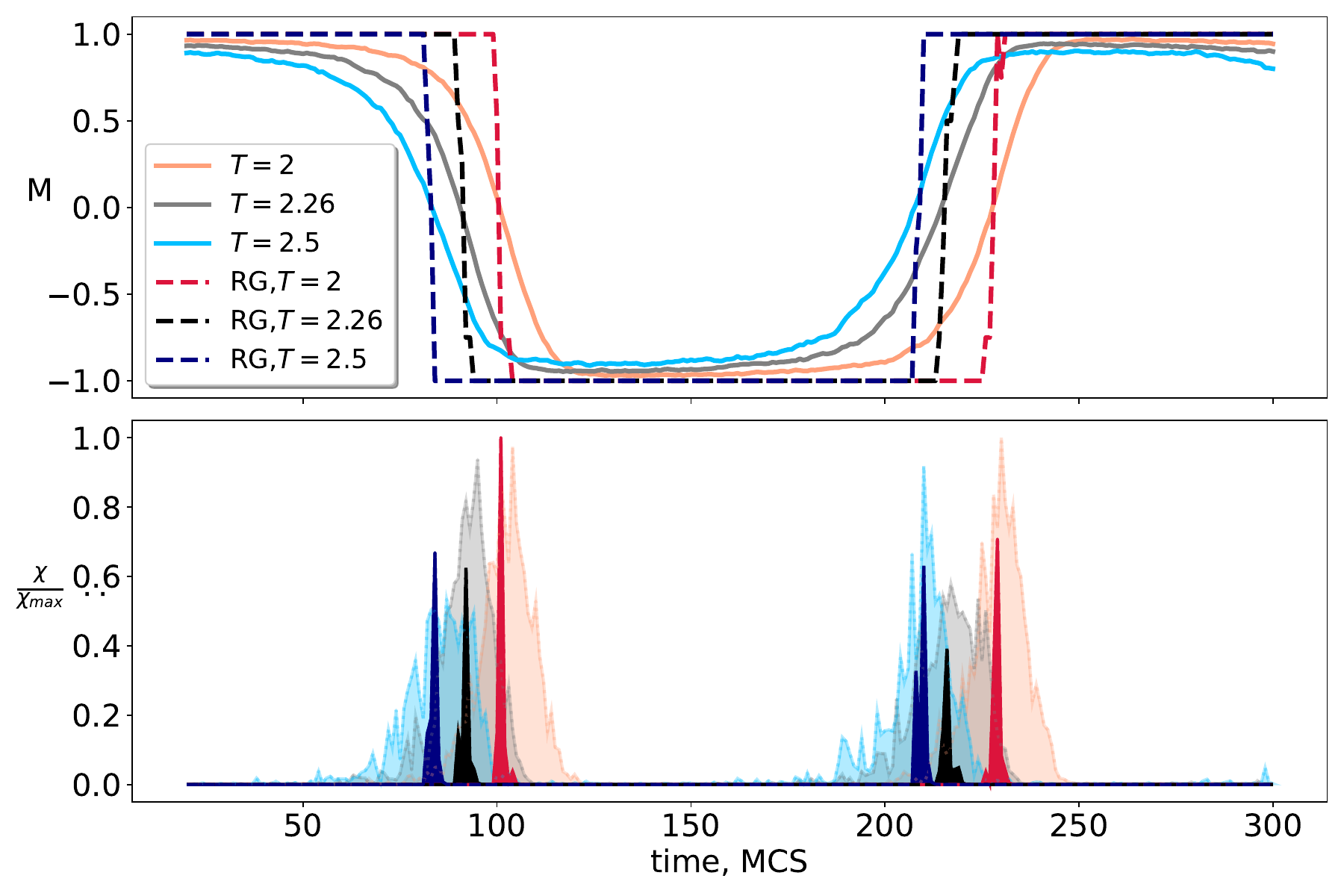}
\caption{\label{figmontecarlo} Monte Carlo Metropolis simulation of 2d Ising model with a time dependent external field for temperatures below critical, critical and above critical: $T_B=2< T_c\approx 2.26 < T_A=2.5$. Short time averages (equations \ref{mediasmoveis}, $r=.9$) for top: Magnetization. Continuous lines for original Ising ($128 \times 128$ lattice, nearest neighbor interaction), dashed lines for system renormalized  by a factor  $b=2$, six times. Bottom: Susceptibility, divided by its maximum value. The susceptibilities of the renormalized system are much larger ($\approx 200$ times) than those of the unrenormalized, and are non zero during a much shorter time, since the transition to a new classification is much faster. }
\end{figure}

We present a third analogy of  a non critical system with large fluctuations.  We use the MCRG algorithm as developed by Swendsen \cite{Swendsen}, for the nearest-neighbor model in 2d in the presence of a slowly changing external uniform field. We use the majority rule with a scale renormalization parameter $b=2$. Figure \ref{figmontecarlo} shows the results of the simulation as  a function of time, measured in Monte Carlo steps, for temperatures above, below and at the critical temperature. Since we are not interested in the well known equilibrium properties of the model, our order parameters are the magnetization $M(t)$ and susceptibility and $\chi(t)$ estimated by a moving average with relaxation:
\begin{eqnarray}
    M(t+1) &=& (1-r )M(t) +r m(t),  \nonumber\\
    \mu_2(t+1) &=& (1-r)\mu_2(t) +r  m(t)^2, \nonumber\\
    \chi(t+1) &=& \mu_2(t+1)-M(t+1)^2, \label{mediasmoveis}
\end{eqnarray}
where $m(t)= \sum_k s_k/N_{RG}$, for a renormalized configuration with  $N_{RG}$ spins. 

In Figure \ref{figmontecarlo} (top) we show the magnetization as the external field changes for three temperatures. The continuous lines are for the unrenormalized model on a 128 $\times$ 128 square lattice and the dashed lines for the majority rule MCRG, scale parameter $b=2$, 6 times renormalized. Two relevant things can be seen; first, the transitions are qualitatively the same, independent of temperature. Second, the magnetizations characterize the macroscopic state that the system would be if the field had been kept constant. Notice that the renormalized magnetization has a much sharper transition of classification than the unrenormalized magnetization. The MCRG magnetization can be seen as the result of a classifier analogous to a convolutional Neural Network, where regions of size $b \times b$ are used to define the "activity" or Kadanoff block spin, in the next layer, i.e the next renormalization stage. 
The "movie" presented to this network is generated by the Monte Carlo procedure. In Figure \ref{figmontecarlo} (bottom) we show, with the same color code, the susceptibilities. The peak at the transitions show that the Monte Carlo dynamics is exploring larger regions of configurations than when the field has a clear sign.  The renormalized susceptibilities are larger in magnitude( $\sim 200$ times) but present much sharper peaks, consistent with the abrupt change in magnetization. the indicator of the classification of microscopic states into  macroscopic state.  This MCRG  uses a predetermined convolutional filter, the simple majority rule. A tunable version \cite{Blote, Swendsen3d}, with probabilistic filters shows, for the Ising 3d nearest neighbor model, a much faster approach to the critical point than with the majority rule. In the language of neural networks this means that adjustments in the weights of the filters may lead to the possibility of the reduction of the number of hidden layers need to converge to a classification.
\section{Conclusions}
We have presented results from simulations of spiking neural networks trained as a classifier. The change in the environmental stimulus induces broad tail  distributions of the truncated correlation of fluctuations of activity in the internal representations in the NN. Before and after this transient period the network settles into a state where these  correlations show very short characteristic time or reduced fluctuations variability. This is similar to results in \cite{Churchland}, where the "stimulus onset quenches neural variability", which they  claim to be a rather general property of the cortex. These broad distributions after input transitions, are found even if we disturb the performance of the classifier by incomplete training or decreased intensity of the input, a simple model of an animal under the effect of anesthesia.  We have defined  avalanches and tried to fit power laws to the empirical distributions of size and duration, for the partially and fully trained NN as well as for the anesthetized model. Crackling noise scaling relations show that, even if there is a critical state, it is not in the directed percolation universality class. But the large increase of $\xi_\tau$ is not a criticality signature. Instead we argue that the role of the external stimulus, analogous to a  magnetic field, is to lead the dynamics from a disappearing basin of attraction to a new one. The crossover from one attractor to another leads to large fluctuations without criticality. To illustrate cases where this occurs, we discussed, first a toy model where a particle  moves  in two dimensions under the influence of an oscillating potential and normal noise. Second, an  infinite range Ising model in an oscillating external field, exactly solvable under the separation of time scales of fast spins and slowly evolving field. Third, we present a MCRG study of an Ising model in 2 dimensions. The interesting thing is that the renormalization of the configurations is mathematically equivalent to a feed-forward convolutional neural network, acting as a classifier. The increases in susceptibility driven by a changing field, are present below, at and above the critical temperature. For the renormalized system the duration  of the susceptibility decreases and its height dramatically increases. These are indications that the renormalization procedure, that filters high frequency spatial Fourier components, is selecting the correct degrees of freedom to classify the thermodynamic state. The experiments with the spiking neural network show that in a changing environment, where the external world stimulus has to drive the dynamics of the network toward a meaningful interpretation, persistent fluctuations of the internal representation will appear as a result of the search for the new basin of attraction associated to a concept that represents the external world. Feed forward neural networks are certainly not models of the brain, but, certainly  the cortex of freely behaving  and even anesthetized animals, are constantly stimulated by changing inputs, which have to be processed by extracting relevant information
 to drive behavioral decisions. This can induce changes in the basin of attraction of the different concepts being identified and this search for the new attractor, generate persistent fluctuations of neuronal activity.

{\bf Acknowledgments} {JHS was supported by a CAPES-Brasil fellowship. This work received support from CNAIPS-USP. The authors declare no conflicts of interest}

\bibliography{referenciasInternalRepresentationSantAnaCaticha}

\end{document}